# Threat Mitigation: The Gravity Tractor
(White Paper 042)

Russell Schweickart, Clark Chapman, Dan Durda, Piet Hut[1]

Summary: The Gravity Tractor (GT) is a fully controlled asteroid deflection concept using the mutual gravity between a robotic spacecraft and an asteroid to slowly accelerate the asteroid in the direction of the "hovering" spacecraft. Based on early warning, provided by ground tracking and orbit prediction, it would be deployed a decade or more prior to a potential impact. Ion engines would be utilized for both the rendezvous with the asteroid and the towing phase. Since the GT does not dock with or otherwise physically contact the asteroid during the deflection process there is no requirement for knowledge of the asteroid's shape, composition, rotation state or other "conventional" characteristics. The GT would first reduce the uncertainty in the orbit of the asteroid via Earth tracking of its radio transponder while station keeping with the asteroid. If, after analysis of the more precise asteroid orbit a deflection is indeed indicated, the GT would "hover" above the surface of the asteroid in the direction of the required acceleration vector for a duration adequate to achieve the desired velocity change. The orbit of the asteroid is continuously monitored throughout the deflection process and the end state is known in real time. The performance envelope for the GT includes most NEOs which experience close gravitational encounters prior to impact and those below 150-200 meters in diameter on a direct Earth impact trajectory.

## I Introduction

The first public presentation of the Gravitational Tractor concept was its publication in the 10 November, 2005 issue of Nature magazine, authored by Drs. Edward Lu and Stanley Love of NASA's Johnson Space Center[2]. As stated in the Nature article the Gravity Tractor (GT) is "…a design concept for a spacecraft that can controllably alter the trajectory of an Earth-threatening asteroid by using gravity as a towline. The spacecraft hovers near the asteroid, with its thrusters angled outwards so that the exhaust does not impinge on the surface. This proposed deflection method is insensitive to the structure, surface properties and rotation state of the asteroid."

In the general case an asteroid deflection mission will be called for when an asteroid 100 meters or more in diameter has been discovered and tracked and determined to have a significant probability of impact with the Earth. The specific timing for the deployment of a deflection mission will depend on many factors, among them the time available prior to impact, the availability of launch opportunities, the calculated probability of impact, and the time required to accomplish a successful deflection. Since the tracking data available for any given asteroid can vary dramatically due to optical and radar tracking limitations, a deflection mission may well need to be deployed prior to a future impact being certain. In some instances (e.g. Apophis) a radio transponder will have to be sent to the asteroid in order to provide adequately accurate and timely information to rationally commit to a deflection[3]. In such instances the GT design can serve the dual role of first determining the precise orbit of the asteroid and therefore the need for deflection, and then, if a deflection is indicated, execute the mission. If, in this circumstance, a deflection is determined not to be required a properly equipped GT spacecraft can instead conduct an asteroid characterization mission.

The technology readiness for a GT design depends strongly on the magnitude of the specific deflection challenge. For many deflection scenarios (albeit presently an unknown percentage of the total) no new technology will be required. In fact in the specific instances of Apophis and 2004VD17 (both of which currently have elevated risk status) as well as an additional statistical subset of deflections, a slight redesign of the already flown Deep Space 1 spacecraft would suffice, using a currently available expendable launch vehicle. For more demanding deflection challenges the GT may still be the best design option, albeit with nuclear electric propulsion (NEP) substituted for solar electric propulsion (SEP). In many cases where the asteroid is in a highly eccentric orbit, NEP, or some equivalent high performance propulsion system will be needed in any event simply to deliver a deflection system of any design to the asteroid.

**II Gravitational Tractor Concept**

The GT concept it extremely simple, depending only on the fundamental gravitational attraction between two masses and the ability of a spacecraft to effectively "hover" above an asteroid's surface. Given the very low gravitational attraction between an asteroid and a spacecraft, this hovering maneuver would be performed using variable thrust ion engines (or the equivalent) to counter the low gravitational force pulling the spacecraft toward the asteroidal surface.

In order to avoid the possibility of the ion engine exhaust plume impinging on the asteroidal surface thereby negating a portion of the thrust and potentially creating a dust cloud, the engines would be angled outward. The specific outward cant of the engines would be determined by a trade-off between the effective loss of thrust due to increasing cant angle and the increased gravitational force (therefore towing force) by minimizing the hovering distance above the surface.

For illustrative purposes this paper assumes a hovering position at 1.5 asteroid radii above the asteroid center of mass. and a plume spreading half angle of 20 degrees. This combination yields an engine cant angle of approximately 60 degrees resulting in an effective thrust of 0.5 the actual combined engine thrust.

This geometry is illustrated in the Nature article and reproduced here as Figure 1.

Of note is the fact that the available towing force is determined entirely by three parameters, the masses of the asteroid and the GT and the hovering distance from the asteroid center of mass. This is specified by the equation;

$$T = GMm/d^2$$

where G is the universal gravitational constant, M is the asteroid mass, m is the GT mass, and d is the distance between the two centers of mass.

The acceleration of the asteroid during the towing maneuver is then simply;

$$\Delta V/sec = Gm/d^2$$

To give a feel for these numbers we can use the asteroid Apophis and a hypothetical 1 metric ton GT as an example. Since Apophis is estimated to be 320 meters in diameter[4] and have a mass of 4.6 x $10^{10}$ kg, the thrust required to hover at 1.5 asteroid radii is 53 millinewtons from each of the two engines (assuming a 60 deg outward cant). For comparison purposes the NSTAR ion engine used on the Deep Space 1 mission generated 92 millinewtons at full power.

The acceleration of the asteroid in this configuration is then $3.7 \times 10^{-5}$ meters/sec/year, a very low acceleration indeed, but sufficient in this case to effect an Apophis deflection in 20 days (see below).

From a design perspective one would want to incorporate multiple ion engines both for redundancy and to potentially provide position control via differential thrusting. Specific spacecraft analysis and design would be required to optimize the configuration for both attitude and position control. While all the essential hardware to execute this hypothetical deflection has already flown, the software for both attitude and position control will have to be developed. Nevertheless there is good precedent in this regard with the generally successful maneuvering of the JAXA Hayabusa mission around the asteroid Itokawa.

**III Asteroid Deflection Requirements**

The applicability of any given deflection technique can be judged only with respect to the requirements presented by the specific NEO impact challenge. These requirements vary greatly based not only on the size/mass of the asteroid and its specific orbit, but also by the time to impact, and most strongly by the question of whether or not the asteroid is headed directly for an Earth impact or for a resonant return keyhole associated with a close Earth encounter prior to the nominal impact.

While asteroid masses of interest may vary by a factor of 1000 (100 meter to 1000 meter diameter) a close encounter with Earth prior to impact can vary the momentum change required for deflection by a factor of 100,000. E.g., in the instance of Apophis the $\Delta V$ required for deflection post 2029 is on the order of $5 \times 10^{-2}$ meters/sec whereas pre 2020 it is as low as $5 \times 10^{-7}$ meters/sec. Clearly there is a very wide range of deflection requirements and therefore a likely need for deflection concepts of widely ranging capability.

Unfortunately there has been no statistical analysis to date of the distribution of potential deflection challenges. Therefore the performance requirements and the probability of need for concepts of various performance levels cannot be authoritatively stated.

In the absence of such an analysis one is left to address specific real examples from the existing database of known asteroids with non-zero probability of Earth impact.

Of the 106 current (23 June 2006) NEOs in the Sentry database[5] with non-zero Earth impact probabilities the two most likely to impact (also the two with the longest tracking arcs) are 99942 Apophis and 2004VD17. The respective impact dates for these asteroids are 2036 and 2102. Both of these NEOs are moderately large and, were it not for the shared characteristic of having close encounters with Earth prior to their impact dates, would be very challenging to deflect. Fortunately, due to the pre-impact close encounters both fall within the range of existing technical capability, *provided* the deflections are initiated early.

A deflection maneuver, if conducted about 2027, would require on the order of $2 \times 10^{-6}$

meters/sec for Apophis and 5 x 10$^{-6}$ meters/sec for 2004VD17. If, however, one were to wait until the asteroid were beyond all gravitational encounters and headed directly for an impact the magnitude of the velocity change would increase by 4.5 and 3.5 orders of magnitude respectively! This extremely powerful influence highlights the need for good tracking information as early as possible and a decision-making process which will enable action to be taken, in many cases, well in advance of the impact date.

It is clear in looking at deflection performance requirements that early, accurate information on the specific requirements over time can make the difference between a relatively modest technical challenge and a challenge which lies beyond the reach of available and/or acceptable technology.

**IV Gravity Tractor Performance**

As stated earlier the performance of a GT (in terms of the accelerating force that can be applied to the asteroid) is dependent on the masses of the asteroid and GT and the distance of the hover above the surface of the asteroid. A more relevant performance parameter for any slowly-acting deflection process is the time it takes to accomplish the required deflection. In the case of the GT the deflection time is then proportional to the square of the hovering distance to the asteroid center of mass and inversely proportional to the mass of the GT itself.

Assuming a hovering "altitude" above the surface equal to half the asteroid mean radius and a very modest 1 metric ton spacecraft a GT would be able to shift the Apophis or 2004VD17 impact points from an assumed direct hit (i.e. centerline of the Earth) to a grazing impact in 20 days and 163 days respectively. If one makes the further assumption that the ion engines used are either the NSTAR engines used and proven on Deep Space 1 (Isp = 3100 sec, maximum thrust = 92 millinewtons at 2.1 kW input power) or the equivalent, the fuel masses required for these deflections are about 10 and 84 kg respectively[6] (see Figure 2).

The orbit of Apophis is quite "Earth-like" with its aphelion never exceeding 1.1 AU. Since the thrust required for the towing maneuver is only 53 millinewtons per engine the Deep Space 1 solar arrays, or equivalent, would be adequate to provide the necessary electrical power for the mission. For plume impingement and redundancy reasons one would want to have at least two engines. Nevertheless, given that the entire Deep Space 1 spacecraft, including scientific experiments, weighed less than 490 kg it appears that a GT for this mission would easily come in below the 1 metric ton level. Ironically it is desirable to actually add mass to the spacecraft up to the limits of the specific launch vehicle selected because the spacecraft mass is what pulls the asteroid.

One further comparison point of interest is that since the Deep Space 1 mission required essentially the identical launch vehicle performance capability as a GT deflection mission to Apophis (DS-1 used the smallest Delta) and the total mission cost for that program was less than $150M (1995-1999 $, including LV) it is clear that an Apophis GT deflection mission would neither be expensive nor technologically challenging.

A GT deflection mission for 2004VD17, on the other hand, while appearing to be just barely feasible would require more detailed analysis with regard to the mass of the spacecraft and the capability of existing expendable launch vehicles to execute this more demanding mission. In particular the orbit of 2004VD17 extends to 2.4 AU at aphelion with a perihelion inside the orbit of Venus. This case illustrates a frequent reality in NEO deflections, i.e. that the effort to get to and rendezvous with a NEO is almost always greater than the effort to deflect it once there.

Nevertheless, using reasonable assumptions it appears that a GT deflection mission to 2004VD17 using both existing spacecraft technology, solar electric propulsion, and an existing launch vehicle is just possible with the primary limitation being launch vehicle capability.

While Apophis and 2004VD17 are both large asteroids (320 and 580 meters diameter respectively) both would be well out of reach of the GT deflection performance capability if it were not for the close gravitational encounters these asteroids experience prior to impact. This stark fact highlights the criticality of determining the statistical probability of such encounters in the general (impacting) NEO population. The analysis further points to the launch vehicle performance limits for any deflection technique where the NEO to be deflected requires a launch vehicle capability greater than a $C_3$ of approximately 80 km$^2$/sec$^2$.

### V Mission Performance Considerations[7]

There are a host of additional considerations which come into play in considering a NEO deflection mission. It is critically important to keep in mind that such a mission falls into a completely different category from the normal scientific research or space exploration mission.

The following comments apply not only to the Gravity Tractor, but to any slowly-acting approach to deflection (serious but different considerations apply to instantaneous deflection or destruction approaches).

A NEO deflection will always be a public safety mission and will, without any doubt, be a major international event with unprecedented media attention. In many, if not most cases, the asteroid's positional error ellipse at the projected time of impact will still exceed the diameter of the Earth at the time when a deflection decision will have to be made, and hence the specific point of impact will not be known. The decision to deflect will often have to be made when the ultimate impact ground zero may be located in any of several countries spread across the face of the planet.

A NEO impact, even one for a specific NEO, is inherently an international affair and the demand for international coordination, if not authorization, will be strong. Given the possibility of failure during the course of deflection, there will be populations and property put at risk that will not have been at risk prior to the deflection operation. Such a potential failure implies a considerable financial liability on the part of the deflecting agency unless indemnified by pre-arrangement with the international community.

These, and other public concerns, argue strongly for there being a very high public confidence in the decision-making process, in the deflection methodology chosen, and in the agency executing the deflection.

With these and other considerations in mind it is critical that any deflection concept must, to be seriously contemplated,
1) be tested and demonstrated prior to use,

2) be capable of providing a precise and timely public announcement of the deflection result (i.e. resulting orbit),
3) be capable of providing assurance to the public that, if created, one or more large fragments do not continue to threaten an impact, and
4) be fully controllable in order that the NEO can be targeted for a specific end state.

Binary systems: Another consideration, not yet well known to the public, is that a significant cohort of NEOs are binary (or multiple) systems. In many cases the secondary is itself large enough to penetrate the atmosphere and cause a threat to life and property. At this time it is thought that 15-20% of NEOs may be binary or multiple systems. In many cases the knowledge of whether or not this situation exists depends on obtaining a radar sighting of the NEO. Given that such radar sightings are rarely available and that the future funding for the powerful Arecibo radar is not assured, this challenge is doubly daunting.

Any deflection technique using an impulsive acceleration will not generally change the orbital track of a secondary. Furthermore unless there is pre-knowledge of a target NEO being a binary system, a separate and perhaps last minute additional mission may have to be mounted to deal with the situation.

Conversely a GT deflection, due to the very low acceleration it imparts to the NEO, will simply cause the secondary to be dragged along with the primary *whether or not* it was known to exist prior to the mission.

One mission vs. two: Another consideration of considerable significance is the fact that a GT mission will fully rendezvous with the NEO at issue, and have aboard a radio transponder. These two facts produce substantial advantages for the GT over any impulsive deflection concept that does not execute a full rendezvous (i.e. match velocity with the NEO), and will provide public confidence in the conduct of the operation unavailable to most other deflection designs.

The criteria on which a NEO deflection decision will be based have not yet been developed. One key factor, however, will be the probability of Earth impact at the time when a mission must be launched to achieve a successful deflection. The probability of impact is directly related to the size of the "target" and inversely proportional to the size of the asteroid's uncertainty ellipse at the time of calculation. For a given deflection under consideration the "target" may be either the Earth itself or a resonant return keyhole associated with a close gravitational encounter preceding the impact. These two different targets can vary in size by many orders of magnitude. For example, in the case of Apophis the effective diameter of the Earth (accounting for gravitational focusing) is 27,600 km. while the width of the 2029 7/6 resonance keyhole is only 600 meters, over 45,000 times smaller! For 2004VD17 the ratio is not as extreme with the 2031 encounter keyhole being approximately 15 km wide and the Earth effective diameter 15,000 km.

In many cases available optical and radar tracking, as well as non-gravitational forces such as the Yarkovsky effect, will result in a residual error ellipse, at the time when a deflection mission must be launched, that will result in considerable uncertainty whether the Earth will be hit at all. For instance, in the Apophis case, which will be a very intensively tracked NEO by 2021 when a deflection mission would have to be launched (17 years of optical tracking and several radar apparitions) the size of the error ellipse will still be so large (~30 km) that even if the asteroid is headed for a direct impact with Earth the calculated probability of impact will be only 1 chance in 125, or less.

For this reason, Steve Chesley of JPL, who has done extensive analysis on this object[8] recommends that if there remains a non-zero impact probability following the 2013 observation of the NEO then a transponder should be deployed to Apophis to further reduce the error ellipse in support of a deflection decision for a 2021 launch.

If the 2013 deployment (assuming a continuing Apophis impact threat) were a Gravity Tractor spacecraft, the entire deflection sequence could be accomplished by a single spacecraft launch since on arrival at the asteroid the GT would first serve as the transponder tracking mission; then if, and only if, the NEO were determined to still be headed toward an impact the spacecraft would shift into position and execute the needed deflection. For any impulsive deflection concept a separate transponder mission would have to be launched and then, if needed, a subsequent mission for the deflection per se.

In a deflection maneuver any mission must plan to deflect the asteroid by a distance (at the time of impact) equal to at least the sum of half the best available residual error ellipse and half the "target" diameter. If the choice for this mission is an impulsive deflection (i.e. any mission which intercepts but does not match velocity with the NEO prior to its deflection operation) the required change in the NEO orbit will dramatically exceed that required for a GT mission (or any concept using a full rendezvous) due to the dramatic reduction in the residual error ellipse as a result of the transponder tracking following arrival at the NEO.

Certainty in results: Finally, and in terms of public acceptance perhaps the most significant consideration of all, is the fact that the transponder aboard the GT spacecraft is available not only for the initial reduction of the remaining error ellipse, but throughout the deflection maneuver and more importantly, at the conclusion of the maneuver. There would be, as a result, no uncertainty as to whether or not the deflection was successful. It would have been tracked continuously from prior to the deflection, throughout the deflection maneuver itself, and at the completion of the deflection. In fact throughout the deflection the maneuver can be extended or otherwise modified in real time, based on ground tracking information. There will be no necessity to rely on assumptions about the response of the asteroid; full knowledge of the progress will be available in real time and adjustments can be made as necessary.

Other keyholes: One further subtle but important distinction remains between the GT/controlled deflection and an impulsive deflection, whether kinetic impact or explosive. That distinction resides in the existence of multiple resonance keyholes populating space nearby the Earth.

Unless a specific final orbit for the NEO to be deflected can be planned and the deflection can be shown to have actually achieved this plan, the public cannot be assured that the NEO itself, or large fragments of it, will not have ended up heading for another keyhole thereby still threatening the Earth. Claims that 'we think it went successfully' will not be adequate.

Until fairly recently the error ellipse for Apophis, despite two years of tracking, contained several resonant return possibilities, including the 8/7 and 15/13 keyholes. Unless a deflection is controlled, i.e. unless it can both target a desired endpoint and guide to it, the result of an impact may well be to simply shift the impact a year or two or ten. Furthermore, unless there is an immediate and precise determination of the post deflection orbit, it may well take considerable time and tracking before the general public can be assured that in fact the asteroid, or a major fragment (in the case where impact or explosion is used), is not still headed for Earth impact.

**VI Conclusions**

The Gravity Tractor deflection concept, while not able to meet all potential deflection challenges, has very substantial advantages over alternative concepts for those applications within its performance range. This performance envelope generally includes most NEOs which experience close gravitational encounters prior to impact and those below 150-200 meters in diameter on a direct Earth impact trajectory

A GT deflection is fully controlled providing an accurate final determination of need for a deflection, the ability to target for and achieve a specific safe final orbit, and precise and immediate knowledge of the final result. No alternative impulsive deflection concept can provide these capabilities.[9]

The Gravity Tractor is also a deflection concept that is available today using not only existing technologies, but for some cases (e.g. for Apophis) existing flown and proven subsystems. The technology readiness level (TRL) for all but the hovering software for these situations sits at TRL = 9, and the software itself, based on the JAXA experience with the Hayabusa mission sits at TRL = 7 or 8.

For NEOs which require a launch vehicle capability beyond that currently available in the inventory (i.e. those in eccentric and/or highly inclined orbits) a development program for high efficiency propulsion will be required. This requirement exists regardless of deflection concept since no system can begin its work until first being delivered to the NEO[10]. Where advanced electric propulsion (NEP or other) is required for the intercept with the NEO the GT will utilize the power source and engines of the spacecraft for the gravity towing operation as well.

Finally it is very critical that neither NASA nor any other agency involved in addressing this challenge underestimate the degree to which the international community, both at the state level and that of the general public, will demand to be involved in and ultimately be satisfied with many of the decisions regarding NEO deflection. Fragmentation of the NEO, uncertainty in the execution and the results, and even nuclear explosions and radiation will be of enormous concern to the world public. Where more certain and benign methods are available to accomplish the deflection such instantaneous but risky approaches will not be acceptable.

The Gravity Tractor, where capable of meeting the deflection challenge, is both technologically and societally the most preferable deflection option.

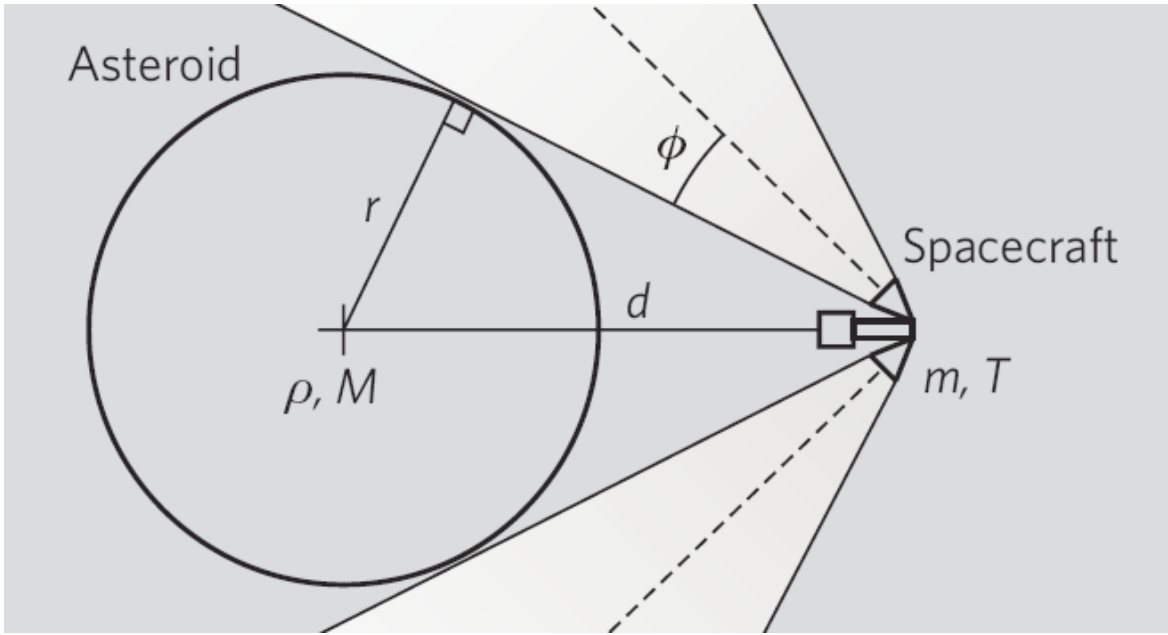

**Figure 1.** The schematic geometry of a Gravity Tractor towing an asteroid. If the distance d between the centers of gravity of the asteroid and GT is equal to 1.5 r and the half-plume angle φ is assumed to be 20 degrees the engines will each have to be canted outward by 60 degrees and each will have to produce thrust equal to the towing force T.

**Gravity Tractor Performance Apophis**

- $M = 4.6 \times 10^{10}$ Kg
- $m = 1 \times 10^{3}$ Kg
- d = 1.5 r = 240 meters
- Ø = 20 degrees
- T = 0.053 Newtons
- $\Delta V = 3.7 \times 10^{-5}$ m/sec/year
- $\Delta V_{req} = 2 \times 10^{-6}$ m/sec
- Deflection $T_{req}$ = 20 days
- $Fuel_{req}$ = ~10 kg

**Gravity Tractor Performance 2004VD17**

- $M = 2.6 \times 10^{11}$ Kg
- $m = 1 \times 10^{3}$ Kg
- d = 1.5 r = 435 meters
- Ø = 20 degrees
- T = 0.092 Newtons
- $\Delta V = 1.1 \times 10^{-5}$ m/sec/year
- $\Delta V_{req} = 5 \times 10^{-6}$ m/sec
- Deflection $T_{req}$ = 163 days
- $Fuel_{req}$ = 83.9 kg

**Figure 2.** Performance figures for a 1 metric ton Gravity Tractor powered by solar electric propulsion in deflecting asteroids 99942 Apophis and 2004VD17, the two NEOs in the current database with the highest probabilities of Earth impact. (M, asteroid mass; m, spacecraft mass; d, hovering distance from asteroid centroid; φ, half angle of ion propulsion plume; T, thrust required to hover; ΔV, deflection acceleration; $\Delta V_{req}$, velocity change required to avoid Earth impact)

[1] Russell L Schweickart, B612 Foundation; Piet Hut, Institute for Advanced Study; Clark Chapman, Dan Durda, Southwest Research Institute.

[2] *Gravitational tractor for towing asteroids*, Edward T. Lu and Stanley G. Love, Nature, Vol. 438, 10 November 2005 (See #12, http://www.b612foundation.org/press/press.html)

[3] *Potential Impact Detection for Near-Earth Asteroids: The Case of 99942 Apophis (2004 MN4)*,Steve Chesley, Asteroids, Comets, Meteors Proceedings, IAU Symposium No. 229, 2005 (See #11, http://www.b612foundation.org/press/press.html)

[4] Impact risk table, JPL, 99942 Apophis, http://neo.jpl.nasa.gov/risks/

[5] Ibid

[6] These fuel figures and deflection times represent a deflection from a centerline Earth impact to a trajectory that just grazes the Earth's limb. Actual mission planning would likely target a specific endpoint several Earth diameters beyond the limb.

[7] This section of the white paper is essentially identical with the comparable section for the Asteroid Tugboat deflection paper since both designs address these mission performance characteristics.

[8] *Potential Impact Detection for Near-Earth Asteroids: The Case of 99942 Apophis (2004 MN4)*,Steve Chesley, Asteroids, Comets, Meteors Proceedings, IAU Symposium No. 229, 2005 (See #11, http://www.b612foundation.org/press/press.html)

[9] There are several conceptual non-impulsive concepts which share these capabilities, but they appear to be far more costly and/or dauntingly complex.

[10] The exceptions to this are those dramatic but risky techniques where a rendezvous is not conducted. Rather a direct impact is planned or a nuclear explosive is set off at precisely the right time as it passes by the NEO. Neither of these schemes can assure the public that there are not large fragments still headed for impact, nor what precisely, in fact, was the outcome of the operation.